\documentclass[aps,showpacs,superscriptadress,preprint]{revtex4}
\usepackage{graphicx}
\usepackage{mathrsfs}

\begin{document}
\title{ Quasinormal modes of gravitational perturbation around some well-known regular black holes}

\author{Chen Wu$^{1}$\footnote{Electronic address: wuchenoffd@gmail.com}} \affiliation{
\small 1. Shanghai Institute of Applied Physics, Chinese Academy of
Sciences, Shanghai 201800, China}

\begin{abstract}
In this paper, the quasinormal modes of gravitational perturbation around some well-known regular black holes were
evaluated by using the  WKB approximation as well as the asymptotic iteration method.  Through numerical calculation, we make a detailed analysis of the gravitational QNM frequencies by varying  the characteristic parameters of the gravitational perturbation and the spacetime charge parameters of the regular black holes. It is found that the imaginary part of quasinormal modes  as  a function of the charge parameter has different  monotonic  behaviors for different black hole spacetimes. Moreover, the asymptotic expressions of gravitational QNMs for $l \gg 1$ are obtained by using the  eikonal limit method. We  demonstrate that the gravitational perturbation is  stable in all these spacetimes.
\end{abstract}

 \maketitle

\section{Introduction}
The investigations concerning the interaction of black holes with various fields
around  give us the possibility to get some information about the physics of black holes. One of these information can be obtained from  quasinormal modes (QNMs) which are characteristic of the background  black hole spacetimes \cite{black hole}.   The concept of QNMs is first formulated by Vishveshwara  in calculations of the scattering of gravitational waves by a black hole \cite{Vishveshwara}, which present complex frequencies whose real part represents the actual frequency of the oscillation and the imaginary part represents the damping. The survey of field perturbation in black hole spacetimes motivated the  extensive numerical and analytical study of QNMs \cite{some papers}. In addition, the properties of QNMs have been studied in the context of the AdS/CFT correspondence \cite{asd/cft} and loop quantum gravity \cite{loop}. Some reviews  where a lot of references to the recent research of QNMs can be found in  Refs. \cite{review}.

 Among several types of field perturbation, gravitational perturbation is considered to be the most important one. The reason is that gravitational QNMs are important in directly identifying  black holes and their gravitational radiation. Many theoretical physicists believe that the  gravitational QNMs is a unique fingerprint in  searching the existence of a black hole. Recently, astrophysical interests in QNMs originated from their relevance in gravitational wave analysis. On September 14th, 2015, two advanced detectors of the Laser Interferometer Gravitational-wave Observatory (LIGO)  made the first direct measurement of gravitational waves  \cite{GW}.  The Advanced LIGO detectors observed a transient gravitational-wave signal determined to be the coalescence of two black holes, launching the era of gravitational wave astronomy. The issue of black hole gravitational stability under perturbations was first addressed by Regge and Wheeler \cite{wheeler} in the fifties of last century. They classified  gravitational perturbations into two types: odd parity and even parity by means of  getting rid of  the angular dependence of the perturbation variables through a tensorial generalization of the spherical harmonics, thus  the calculation of  gravitational perturbation was greatly simplified.  The Regge-Wheeler formalism was later extended to the case of  static black holes in four dimensions \cite{G QNMs 4D-1,G QNMs 4D-2, G QNMs 4D-3} and  higher dimensions \cite{G QNMs HD-1}, and even to the case of rotating black holes \cite{G QNMs Rotating-1,  G QNMs Rotating-2}. One can find a complete description of black hole perturbation theory in the book by Chandrasekhar \cite{G QNMs book}.

On the other hand, the problem of  understanding how to avoid singularities in black hole spacetimes is important in general relativity.
In 1968, a ``regular'' black hole without a singularity  was constructed by Bardeen \cite{Bardeen}.  This regular black hole spacetime lacked a consistent physical interpretation  until Ay\'{o}n-Beato and his coworkers \cite{ABG2000} obtained this black hole solution by describing it as the gravitational field of nonlinear magnetic monopole with a mass $M$ and a charge $q$ in 2000.

 The Bardeen solution has motivated deeper works about  singularity avoidance may be realized generally. Several other researchers paid attention to theories of gravity coupled to nonlinear electrodynamics, and  proposed other solutions in different contexts.  Some solutions that are relevant to this work are analyzed  in Refs. \cite{Hayward, Bronnikov, Dymnikova, ABG, new1} and Refs. \cite{Bardeen, ABG2000}.

Recently, there are several interesting works concerning the regular black hole. Eiroa and Sendra  have investigated
gravitational lensing of the regular black hole spacetime \cite{eiroa1}.  In Ref. \cite{f(r)},  exact solutions of  spherically symmetric spacetimes are proposed in $f(R)$ modified theories of gravity coupled to nonlinear electrodynamics. The dynamical stability of black hole solutions in self-gravitating nonlinear electrodynamics with respect to  linear gravitational fluctuation has been studied in \cite{stability regularBH}.
Fernando and Correa have studied  QNMs spectrum of the scalar field of the regular black hole for various values of the perturbation parameters \cite{Fernando}.
They have also used the unstable null geodesics of the black hole to compute the scalar QNMs in the eikonal limit. Further research about
 the QNMs of neutral and charged scalar field perturbations on the regular black hole spacetime in a variety of models was carried out by Flachi and Lemos \cite{Flachi}. Massless and massive Dirac QNMs were studied in  the regular black hole spacetime by using the WKB approach in Ref. \cite{Dirac}.

In this paper, we concentrate on the behavior of the  gravitational perturbation in the regular black hole spacetimes mentioned above.   The wavelike perturbation equation, with an effective potential, can be solved by several methods, such as integration of the wavelike equations, the monodromy method, fit and interpolation approaches,  the continued fraction method,  the Mashhoon method, the WKB approximation method, the asymptotic iteration method and so on \cite{QNM method}. We calculate the gravitational QNMs by  using the 3th order WKB method as well as the asymptotic iteration method.

 The rest of the paper is organized as follows. In section 2, we gives brief description of some well-known regular black hole spacetimes.  The perturbative equation of the  gravitational perturbation in given backgrounds is reduced to the Schr\"{o}dinger-like wave equation in section 3. The next section is devoted to  the numerical calculations of the gravitational QNMs in given spacetimes  by using the 3th order WKB approximation and the asymptotic iteration method. The eikonal limit for the gravitational frequencies is also presented. The conclusions are given in last section.

\section{The basic equations and numerical results }
In this section, we will first give a brief  introduction to the regular  black holes. In order to obtain the regular solutions, the typical  action to include the nonlinear electrodynamic term is,
\begin{equation}
S = \frac{1}{16 \pi G} \int d^4x \sqrt{-g} \left( R -  \mathscr{L}(F) \right),
\end{equation}
where $g$ is the determinant of the black hole metric, $G$ is the gravitational constant, $R$ is the scalar curvature, and $\mathscr{L}(F)$ represents  the Lagrangian of the  nonlinear electrodynamics with   $F = \frac{1}{4}F_{\mu \nu} F^{\mu \nu}$, where $ F_{\mu \nu} =  \bigtriangledown_{\mu} A_{ \nu} - \bigtriangledown_{\nu} A_{ \mu} $ is the electromagnetic field strength.

The general line element for spherically symmetric regular black hole solutions  can be described as
\begin{equation}
ds^2 = - f(r) dt^2 + \frac{dr^2}{f(r)} +r^2 d\theta^2 +  r^2\sin^2\theta\,d\phi^2~,
\label{rbhm}
\end{equation}
where $(t,r,\theta,\phi)$
are the usual space-time spherical coordinates, and  specific choices of  the lapse function $f(r)$ distinguish between the different spacetimes.

In Ref. \cite{Bardeen},  the lapse function $f(r)$ for the Bardeen black hole is determined by the formula
\begin{equation}\label{bardeen metric}
f(r) = 1-\frac{2Mr^2}{(r^2+ q^2)^{3/2}}\,,
\end{equation}
where $q$ and $M$ are the magnetic charge and the mass of the magnetic monopole.
 Ay\'{o}n-Beato and his coworker interpreted this black hole as the gravitational field of a magnetic monopole arising from nonlinear electrodynamics \cite{ABG2000}. The Lagrangian of the specific  nonlinear electrodynamics is   given by
$\mathscr{L}(F)=(3M/|q|^3)(\sqrt{2q^2F}/(1+\sqrt{2q^2F}))^{5/2}$. The spacetime in Eq. \ref{bardeen metric} has horizons only if
$ |q| \leq \frac{ 4 }{ 3 \sqrt{3}}$.  For $ q = \frac{ 4 }{ 3 \sqrt{3}}$, there are degenerate horizons. For $ q > \frac{ 4 }{ 3 \sqrt{3}}$, there are no horizons.

In 2006, Hayward found new regular black hole spacetime that has center flatness and is quiet similar to the physical insight of the Bardeen one \cite{Hayward}.
The simple regular black hole implies a specific matter energy-momentum tensor that is de Sitter at the core and vanishes at large distances $r \rightarrow \infty$. The function $f(r)$ for the Hayward black hole also takes a simple form
\begin{equation} \label{hayward}
f(r) = 1-\frac{2Mr^2}{r^3+2\alpha^2}\,
 \end{equation}
with $\alpha =$ Const. As like the Bardeen black hole, the Eq. (\ref{hayward}) can also have zero, one, or two horizons depending on the relative values of $M$ and $\alpha$.

  Through introducing the Lagrangian for nonlinear electrodynamics to first order, the regular black hole is also constructed originally by Bronnikov \cite{Bronnikov}. The lapse function for this black hole is given as
  \begin{equation} \label{Bronnikov}
  f=1-{2M\over r}\left(1- \mbox{tanh}{r_0\over r}\right),
  \end{equation}
where the parameter $r_0$  is a length scale related to the electric charge.

Then Dymnikova \cite{Dymnikova} put forward an exact, regular spherically symmetric, charged black hole solution by using the idea proposed by Bronnikov \cite{Bronnikov}. This solution is constructed from a nonlinear electrodynamic theory with a Hamiltonian-like function (for more details, please see Ref. \cite{Dymnikova}). The lapse function for Dymnikova's solution is given as
\begin{equation}
f(r) = 1- \frac{4M}{ \pi r} \left(\mbox{tan}^{-1}{r\over r_0} -
{r r_0 \over r^2+r_0^2}\right)~.
\label{Dymnikova}
\end{equation}
The parameter $r_0$ in the solution (\ref{Dymnikova}) is a length scale defined as $r_0=\pi q^2/ (8M)$, where $M$ and $q$ is the
 total mass  and  the charge.

Another famous regular black hole is the one presented by Ay\'on-Beato and Garc\'ia in Ref.~\cite{ABG}. To find this regular black hole solution, they took the nonlinear electric field  as a source of charge  for the solution of Maxwell field equations.      Its lapse function $f(r)$ is given as
\begin{equation}
f= 1-{2Mr^2\over (r^2+q^2)^{3/2}}+{q^2r^2\over (r^2+q^2)^2}~,
\label{ABG1998}
\end{equation}
where $M$ and $q$ are the total mass and charge. This solution can be obtained from a nonlinear electrodynamics with Lagrangian density
$ {\mathscr L}(F) = {X^2\over -2q^2}{1-8X-3X^2 \over \left(1-X\right)^4}
-{3M\over 2 q^3 }{X^{5/2}\left(3-2X\right) \over \left(1-X\right)^{7/2}}~, $ where $X=\sqrt{-2q^2F}$.

Recently Balart and Vagenas \cite{new1} built  charged regular black holes in the framework of Einstein$-$nonlinear electrodynamics theory. They constructed the
general lapse function for mass distribution functions that are inspired by continuous probability distributions. In this paper, we consider two examples of black hole solutions employing their methodology. One  metric function is of the form
\begin{equation}
f= 1- \frac{2M}{r} e^{-\frac{q^2}{2Mr}}~.
\label{new1-1}
\end{equation}
 The other metric function is  written as
\begin{equation}
f= 1- \frac{2M}{r} \frac{2} { \text{exp}( {\frac{q^2}{Mr}}) +1 }~.
\label{new1-2}
\end{equation}
In two metric functions $M$ and $q$ are associated with total mass and charge, respectively.

All the black hole solutions used in this paper are summarized in Table \ref{metrics}. The mass of black holes is normalized to 1.  It is noteworthy that  in all the mentioned black holes the lapse function $f(r)$ can have zero, one, or two horizons depending on the  value of charge parameters. In Table \ref{metrics}, we list the extreme charge parameters for which the inner horizon  and outer horizon coincide. To illustrate the behavior of the lapse function more clearly, we show the lapse function as a function of $r$ for three values of the charge parameter in Fig. 1. Noting that the plots of all lapse functions for regular black holes under consideration  show similar behaviors, we take the plot of the lapse function for the Bardeen regular black hole as an example.

\begin{table}[hbt]\centering\caption{Summary of some well-known regular black holes in the paper. For more details, one can refer to the Refs. \cite{Bardeen, ABG2000, Hayward, Bronnikov, Dymnikova, ABG, new1} listed in the second section. In this work, the mass of black holes is normalized to 1.}
\begin{tabular*}{16.5cm}{*{4}{c @{\extracolsep\fill}}}
\hline \hline
Lapse function & Reference  & Extremal condition & Originator    \\ \hline
$f(r) = 1-\frac{2Mr^2}{(r^2+q^2)^{3/2}}$ & \cite{Bardeen} & $q \approx 0.77$  & Bardeen       \\
$f=1-{2Mr^2\over r^3+2\alpha^2}$ & \cite{Hayward} &$\alpha \approx 1.06 $   & Hayward \\
$f=1-{2M\over r}\left(1- \mbox{tanh}{r_0\over r}\right)$ &\cite{Bronnikov} & $r_0 \approx 0.55$   & Bronnikov  \\
$f=1-{4M\over \pi r}\left(\mbox{tan}^{-1}{r\over r_0}- {rr_0\over r^2+r_0^2}\right)$ & \cite{Dymnikova} &$r_0 = 0.45$   &Dymnikova  \\
$f= 1-{2Mr^2\over (r^2+q^2)^{3/2}}+{q^2r^2\over (r^2+q^2)^2}$ &\cite{ABG}  &$q \approx 0.63 $    & Ay\'on-Beato and Garc\'ia \\
$f=1 - \frac{2M}{ r} e^{-q^2/{2mr} }$     &\cite{new1}  &$q \approx 1.21$   & Balart and Vagenas \\
$f=1 - \frac{2M}{ r} \frac{2}{e^{q^2/{mr}}+1 }$ &\cite{new1}  &$q \approx 1.06$    & Balart and Vagenas \\  \hline \hline
\end{tabular*} \label{metrics}
\end{table}

\section{ Perturbation equation for the gravitational field}

The investigation of black hole perturbations  was first carried out by Regge
 and Wheeler \cite{wheeler} for the  odd  parity type of the spherical harmonics  and was
 extended to the even  parity type by Zerilli \cite{zerilli}.

 We indicate the background metric with $g_{\mu\nu}$ and the
 perturbation in it with $h_{\mu\nu}$. The perturbation $h_{\mu\nu}$
 is  very small compared with $g_{\mu\nu}$. The
 $R_{\mu\nu}$ can be expressed from $g_{\mu\nu}$, and $R_{\mu\nu}+\delta
 R_{\mu\nu}$ from $g_{\mu\nu}+h_{\mu\nu}$. $\delta
 R_{\mu\nu}$ can be  calculated from the  form \cite{book}
 \begin{equation}\label{DeltaR}
 \delta R_{\mu\nu}=-\delta \Gamma^{\beta}_{\mu\nu;\>\beta}+\delta
\Gamma^{\beta}_{\mu\beta;\>\nu},
\end{equation}
 where \begin{equation}\label{eq:3}
 \delta\Gamma^{\alpha}_{\beta\gamma}=\frac{1}{2}g^{\alpha\nu}(h_{\beta\nu;\>\gamma}+h_{\gamma\nu;\>\beta}-h_{\beta\gamma;\>\nu}).
 \end{equation}

The canonical form for the perturbations in the Regge-Wheeler gauge
is given as \cite{wheeler}

\begin{equation}\label{h}
 h_{\mu\nu}=
 \begin{array}{|cccc|}
  0&0&0&h_{0}(r)
 \\0&0&0&h_{1}(r)
 \\0&0&0&0
 \\h_{0}(r)&h_{1}(r)&0&0
 \end{array}\exp(-i\omega t) \sin\theta  \frac{\partial Y_{L0}}{ \partial\theta}
 \end{equation}

Substituting Eq. (\ref{h}) into Eq. (\ref{DeltaR}), we  get
 \begin{equation}
 \frac{1}{f(r)}i\omega h_0(r)- \frac{d}{dr}f(r)h_1(r )= 0,  \;\; \text{from} \;\; \delta R_{23}=0
 \end{equation}
 \begin{equation}
 \frac{1}{f(r)} i\omega \left(i\omega h_1(r) + \frac{d}{dr}h_0(r) - \frac{2}{r}h_0(r) \right) + \frac{1}{r^2}(l(l+1)-2)h_1(r) = 0,  \;\; \text{from} \;\; \delta R_{13}=0
 \end{equation}
Defining
 \begin{equation}
 \phi = f(r)h_1(r)/r,
 \end{equation}
 and eliminating the $h_0(r)$ we  get
\begin{equation}
  (\frac{d^{2}}{dr_{*}^{2}}+\omega^{2})\Phi(r)=  V(r) \Phi(r), \label{wave equation}
 \end{equation}
 where $r_*$ is the tortoise coordinate and
 \begin{equation}
  V(r) = f(r)\left( \frac{l(l+1)}{r^2} - 2\frac{1-f(r)}{r^2} - \frac{1}{r}\frac{df(r)}{dr}  \right). \label{effective potential}
 \end{equation}

As mentioned before, the complex $\omega$ values are written as   $\omega = \text{Re}(\omega) + i \text{Im}(\omega)$. The effective potential $V(r)$ of wavelike perturbation equation for the   black hole spacetimes is plotted to display how it changes with the  angular harmonic index  $l$  in Fig. 2. One can find from Fig. 2 that the  angular harmonic index $l$ increases the height of the potential barrier governed by the effective potential.

\section{Numerical methods and numerical results}

Now we report the QNM frequencies of the gravitational perturbation in the regular black hole spacetimes. In order to calculate QNMs, boundary conditions are imposed  for the wavelike perturbation equation.  The boundary condition at the horizon is for the solution to be purely ingoing wave, while the other boundary condition at spatial infinity is     such that the wave has to be purely outgoing one.  Therefore one can write the boundary conditions as
\begin{equation}
\Phi(r) \sim e^{-i\omega r_*}, \;\;\;\; \text{as}  \;\;\;\; r_* \rightarrow -\infty,
\end{equation}
\begin{equation}
\Phi(r) \sim e^{ i\omega r_*},\;\;\;\; \text{as}  \;\;\;\; r_* \rightarrow +\infty.
\end{equation}

\textbf{WKB method.} The Schr\"{o}dinger-like wave equation (\ref{wave equation}) with the effective potential (\ref{effective potential}) containing the lapse function f(r) related to the regular black holes is not solvable analytically. Many numerical methods are developed to compute QNMs of various black hole spacetimes in the literature.
One of these standard methods is the WKB approximative method that was applied for the first time by Schutz and Will \cite{WKB}.
  Iyer and his coworkers developed the WKB method up to third order \cite{WKB3} and later, Konoplya developed it up to sixth order \cite{WKB6}. This semianalytic method has been applied extensively in numerous black hole spacetime cases, which has been proved to be accurate up to around one percent for  the real and the imaginary parts of the quasinormal frequencies for low-lying modes with $n<l$, where $n$ is the mode number and $l$ is the angular momentum quantum number. In this paper, the third order WKB formalism is applied since the 6th order WKB method consumes  too much CPU power for some cases of regular black holes.

The 3th order formalism of the WKB approximation  presented in the paper \cite{WKB3} has formula
\begin{equation}
\frac{ \omega^2 - V_0}{ \sqrt{ - 2 V_0^{''}}} - L_2 - L_3  = n + \frac{1}{2}.
\end{equation}
Here, $V_0$ and $V_0^{''}$ are the maximum potential and the second derivative of the potential evaluated at the maximum potential, $n$ is the node number, and $L_n$ represents the n-th order correction.  The formulae for $L_2$ and $L_3$ are given in \cite{WKB3}.

\textbf{The eikonal limit.} It is well-know   that the WKB method works with high accuracy for large values of the multipole quantum number. For $l \gg 1$, frequency of QNMs of gravitational perturbation field  can be found analytically by using the first order WKB approach. Since the lapse function is complicated in  any regular black hole metric,  a more convenient method is used to obtain the  asymptotic form of frequency of QNMs of gravitational perturbation field.
For $l \gg 1$, the effective potential $V(r)$ at its maximum  has  a asymptotic form: $V(r_{Max}) \propto l(l+1)$. Then the QNM frequency takes the form
\begin{eqnarray}\label{limit1}
\lim_{l \to\infty} \omega \approx c_1(l+\frac{1}{2})-i c_2(n+\frac{1}{2})\ .
\end{eqnarray}
We can readily fix $c_1$ and $c_2$  by setting $l \gg 1$ in numerical calculations. In Table. \ref{eikonal}, we list different $c_1$ and $c_2$ for several regular black hole spacetimes.

\begin{table}[hbt]\centering\caption{The constant $c_1$ and $c_2$ in the eikonal limit for given regular black hole spacetimes.}
\begin{tabular*}{16.5cm}{*{5}{c @{\extracolsep\fill}}}
\hline \hline
Lapse function & Ref. &Parameter &$c_1$ &$c_2$   \\ \hline
$f(r) = 1-\frac{2Mr^2}{(r^2+\alpha^2)^{3/2}}$ & \cite{Bardeen} & $\alpha$ = 0.6 & 0.2066 & 0.1796      \\
$f=1-{2Mr^2\over r^3+2\alpha^2}$ & \cite{Hayward} &$\alpha = 0.7$   &0.2009 & 0.1683\\
$f=1-{2M\over r}\left(1- \mbox{tanh}{r_0\over r}\right)$ &\cite{Bronnikov} &$r_0 = 0.4$   &0.2301 & 0.1947  \\
$f=1-{4M\over \pi r}\left(\mbox{tan}^{-1}{r\over r_0}- {rr_0\over r^2+r_0^2}\right)$ & \cite{Dymnikova} &$r_0 = 0.4$   &0.2491 & 0.1867 \\
$f= 1-{2Mr^2\over (r^2+q^2)^{3/2}}+{q^2r^2\over (r^2+q^2)^2}$ &\cite{ABG}  &$q = 0.6$   &0.2267 & 0.1712 \\
$f=1 - \frac{2M}{ r} e^{-q^2/{2mr} }$     &\cite{new1}  &$q = 0.6$   &0.2053 & 0.1961 \\
$f=1 - \frac{2M}{ r} \frac{2}{e^{q^2/{mr}}+1 }$ &\cite{new1}  &$q = 0.6$   &0.4493 & 0.3917\\  \hline \hline
\end{tabular*} \label{eikonal}
\end{table}

\textbf{The asymptotic iteration method.} The asymptotic iteration method (AIM) was first applied to solve  the second order differential equations \cite{Ciftci}. This new method was then used to obtain the QNM frequencies of field perturbation in Schwarzschild black hole spacetime ~\cite{AIM2010}.

Let's consider a  second order differential equation of the form
\begin{equation}\label{2nd Eq}
\chi''=\lambda_{0}(x)\chi'+s_{0}(x)\chi,
\end{equation}
where $\lambda_{0}(x)$ and $s_{0}(x)$ are   well defined functions and sufficiently smooth. Differentiating the equation above with respect to $x$  leads to
\begin{equation}
\chi'''=\lambda_{1}(x)\chi'+s_{1}(x)\chi,
\end{equation}
 where  the new two coefficients are  $\lambda_{1} (x) =\lambda_{0}'+s_{0}+\lambda_{0}^{2}$ and $s_{1}(x)=s_{0}'+s_{0}\lambda_{0}.$

 Using this process iteratively, we can get the $(n+2)-$th  derivative of $\chi(x)$ with respect to $x$ as
\begin{equation}
\chi^{(n+2)}=\lambda_{n}(x)\chi'+s_{n}(x)\chi,
\end{equation}
 where  the new coefficients $\lambda_{n}(x)$ and $s_{n}(x)$ are associated with the older ones through the following relation
\begin{equation}
\lambda_{n}(x)=\lambda'_{n-1}+s_{n-1}+\lambda_{0}\lambda_{n-1}, \;\;\;\; s_{n}(x)=s'_{n-1}+s_{0}\lambda_{n-1}.\label{iteration}
\end{equation}

 For sufficiently large values of $n$, the asymptotic concept of the AIM method is introduced by ~\cite{AIM2012},
\begin{equation}
\frac{s_{n}(x)}{\lambda_{n}(x)}=\frac{s_{n-1}(x)}{\lambda_{n-1}(x)} = \text{Constant}. \label{Quantum condition}
\end{equation}
The perturbation frequencies can be obtained from  the above ``quantization condition''. However this procedure has a difficulty in that the process of taking the derivative of $\lambda_{n}(x)$ and $s_{n}(x)$ terms terms of the previous iteration at each step can consume much time and  affect the numerical precision of calculations. To  overcome these drawbacks,   $\lambda_{n}(x)$ and $s_{n}(x)$ are expanded  in Taylor series around the point $x'$ at which the
AIM method is performed \cite{AIM2010},
\begin{eqnarray}
\lambda_{n}(x')=\sum_{i=0}^{\infty}c_{n}^{i}(x- x')^{i} & , & s_{n}(x')=\sum_{i=0}^{\infty}d_{n}^{i}(x-x')^{i},
\end{eqnarray}
 where $c_{n}^{i}$ and $d_{n}^{i}$ are the $i$-th Taylor coefficients of $\lambda_{n}(x')$ and $s_{n}(x')$, respectively. Substitution of above equations into Eq. (\ref{iteration}) leads to a set of recursion relations for the Taylor coefficients as
\begin{eqnarray}
c^{i}_{n}=  \sum_{k=0}^{i}c_{0}^{k}c_{n-1}^{i-k} + (i+1)c_{n-1}^{i+1}+ d^i_{n-1} & , & d_n^i= \sum_{k=0}^{i}d_{0}^{k}c_{n-1}^{i-k} + (i+1)d^{i+1}_{n-1}.
\label{recursion}
\end{eqnarray}
After applying the  recursion relations (\ref{recursion}) in Eq. (\ref{Quantum condition}), the quantization condition then can  be expressed   as
\begin{equation}
d_{n}^{0}c_{n-1}^{0}- c_{n}^{0} d_{n-1}^{0}=0,
\end{equation}
 which can be employed to calculate the QNMs of  a black hole. Both the accuracy and efficiency of the AIM method are greatly improved  without any derivative computation \cite{AIM2010}.

First, we calculate gravitational QNM frequencies by varying  the charge of the Bardeen regular black hole. We  also calculate the gravitational QNM frequencies for the Reissner-Nordstr$\ddot{\text{o}}$m black hole with the same  charge for comparison as shown in Table \ref{compare}.  Here, the node number $ n=0$ and the angular momentum quantum number $l = 2$. From Table. \ref{compare}, it is shown clearly that the real part $\text{Re}(\omega)$ of the gravitational QNMs  increases when charge parameter $q$ increases for the two black hole spacetimes. And the imaginary value $\text{Im}(\omega)$ of QNMs decreases for the Bardeen black hole spacetime with  $q$, while for the Reissner-Nordstr$\ddot{\text{o}}$m black hole spacetime, the imaginary value  arrives a maximum 0.0909 around $q = 0.76$.

 \begin{table}[tbh]
\centering\caption{The gravitational QNMs in the Bardeen black hole spacetime  as a function of the \\ charge parameter.
As comparison, we also  calculate the QNMs in the Reissner-Nordstr$\ddot{\text{o}}$m black hole spacetime. The numerical results are computed by the AIM method. Here,  $ n=0$ and  $l = 2$.}
\begin{tabular*}{16.5cm}{*{4}{c @{\extracolsep\fill}}}
\hline \hline
 q & $\omega$ (RN black hole)  &  $\omega$ (Bardeen black hole) \\ \hline

0.1 & 0.3743 $-$ 0.0890i  & 0.3743 $-$ 0.0889i \\ \hline

0.2 & 0.3763 $-$ 0.0892i & 0.3765 $-$ 0.0885i \\ \hline

0.3 &  0.3797 $-$ 0.0894i & 0.3801 $-$ 0.0879i \\ \hline

0.4 &  0.3847 $-$ 0.0898i & 0.3856 $-$ 0.0870i \\ \hline

0.5  & 0.3916 $-$ 0.0902i & 0.3933 $-$ 0.0855i \\ \hline

0.6  & 0.4008 $-$ 0.0906i & 0.4039 $-$ 0.0829i \\ \hline

0.7  & 0.4130  $-$ 0.0909i & 0.4189 $-$ 0.0782i \\ \hline

0.76  & 0.4244  $-$ 0.0909i  & 0.4309 $-$ 0.0728i   \\ \hline

0.8  & 0.4297  $-$ 0.0908i  & **  \\ \hline

0.85  & 0.4404 $-$ 0.0903i  & ** \\ \hline\hline
\end{tabular*} \label{compare}
\end{table}

Results for the quasinormal frequencies for gravitational perturbations in regular black hole spacetimes under consideration are tabulated, in Table \ref{tableBardeen} for the solution given in Refs. \cite{Bardeen, ABG2000}, in Table \ref{tableHayward} for the solution given in Ref. \cite{Hayward}, in Table \ref{tableBronnikov} for the solution given in Ref. \cite{Bronnikov}, in Table \ref{tableDymnikova} for the solution given in Ref. \cite{Dymnikova}, in Table \ref{tableABG} for the solution given in Ref. \cite{ABG}, in Table \ref{tablenew1-1} for the solution given in Ref. \cite{new1}, and in Table \ref{tablenew1-2} for the solution given in Ref. \cite{new1}.  The values of the quasinormal frequencies  listed in these tables are computed by the third order WKB method (without parenthesis) and the AIM method  (with parenthesis), respectively. It has been shown that  the linear perturbative gravitational field are stable around all of the considered regular black holes.

We can apply the sixth order WKB approximation to check the convergence of the WKB approximation.   In Fig. 3, The real   and imaginary  parts of the QNMs from gravitational perturbations for the $l = 2, n = 0$ mode for the model of Bardeen \cite{Bardeen} are presented when higher order terms in the WKB approximation are included in the computation. It shows that the accuracy of the third order WKB method is reliable.

The above calculations have shown that increasing of the spacetime charge parameter implies monotonic increasing of the real part of quasinormal frequency. However, the imaginary part of quasinormal frequency  as  a function of the charge parameter has different monotonic behaviors for different black hole spacetimes.

 For the Bardeen spacetime, the Hayward spacetime and the solution of Ref. \cite{ABG}, increasing of the spacetime charge parameter implies monotonic decreasing of the imaginary part of QNM frequency; For the two solutions of Ref. \cite{new1}, the imaginary part of QNM frequency increases when the charge parameter increases; However, for the two solutions of  Ref. \cite{Bronnikov, Dymnikova}, there exists a maximum of the imaginary part in the charge parameter interval.

 \begin{table}[tbh] \label{tableBardeen}
\centering\caption{ QNMs for gravitational perturbations for $q=$ 0.1, 0.3, 0.6  for the model of Refs. \cite{Bardeen, ABG2000}. }
\begin{tabular*}{16.5cm}{*{5}{c @{\extracolsep\fill}}}
\hline \hline
$l$ &n  & $q=0.1$   &  $q=0.3$ &  $q=0.6$ \\ \hline

2 & 0  & 0.3738 $-$ 0.0891i  & 0.3795 $-$ 0.0881i & 0.4030 $-$ 0.0828i\\ \hline

 &   & (0.3743 $-$ 0.0889i)  & (0.3801 $-$ 0.0879i) & (0.4039 $-$ 0.0829i)\\ \hline

2 & 1 & 0.3468 $-$ 0.2745i & 0.3539 $-$ 0.2712i & 0.3820 $-$ 0.2536i\\ \hline

&   & (0.3476 $-$ 0.2736i)  & (0.3552 $-$ 0.2705i) & (0.3850 $-$ 0.2534i)\\ \hline

3 &0 &  0.6003 $-$ 0.0926i & 0.6091 $-$ 0.0917i  & 0.6455 $-$ 0.0866i \\ \hline

&   & (0.6005 $-$ 0.0926i)  & (0.6093 $-$ 0.0917i) & (0.6457 $-$ 0.0866i)\\ \hline

3 &1 &  0.5835 $-$ 0.2811i & 0.5933 $-$ 0.2782i  & 0.6328 $-$ 0.2620i \\ \hline

&   & (0.5838 $-$ 0.0809i)  & (0.5936 $-$ 0.2781i) & (0.6335 $-$ 0.2619i)\\ \hline

3 &2 & 0.5546 $-$ 0.4761i & 0.5659 $-$ 0.4708i & 0.6102 $-$ 0.4421i \\ \hline

&   & (0.5531 $-$ 0.4785i)  & (0.5648 $-$ 0.4730i) & (0.6107 $-$ 0.4429i)\\
\hline\hline
\end{tabular*}
\end{table}

 \begin{table}[tbh]
\centering\caption{ QNMs for gravitational perturbations for $\alpha = 0.1, 0.4, 0.7$  for the model of Refs. \cite{Hayward}. }
\begin{tabular*}{16.5cm}{*{5}{c @{\extracolsep\fill}}}
\hline \hline
$l$ &n  & $\alpha=0.1$   &  $\alpha=0.4$ &  $\alpha=0.7$ \\ \hline

2 & 0  & 0.3735 $-$ 0.0891i  & 0.3782 $-$ 0.0864i & 0.3903 $-$ 0.0776i\\ \hline

 &   & (0.3740 $-$ 0.0888i)  & (0.3791 $-$ 0.0864i) & (0.3920 $-$ 0.0781i)\\ \hline

2 & 1 & 0.3464 $-$ 0.2744i & 0.3521 $-$ 0.2659i & 0.3619 $-$ 0.2378i\\ \hline

&   & (0.3472 $-$ 0.2734i)  & (0.3549 $-$ 0.2652i) & (0.3696 $-$ 0.2374i)\\ \hline

3 &0 &  0.5997 $-$ 0.0926i & 0.6072 $-$ 0.0902i  & 0.6470 $-$ 0.0820i \\ \hline

&   & (0.5999 $-$ 0.0926i)  & (0.6075 $-$ 0.0902i) & (0.6274 $-$ 0.0820i)\\ \hline

3 &1 &  0.5829 $-$ 0.2810i & 0.5914 $-$ 0.2734i  & 0.6108 $-$ 0.2478i \\ \hline

&   & (0.5832 $-$ 0.2808i)  & (0.5922 $-$ 0.2733i) & (0.6126 $-$ 0.2477i)\\ \hline

3 &2 & 0.5539 $-$ 0.4759i & 0.5638 $-$ 0.4626i & 0.5806 $-$ 0.4182i \\ \hline

&   & (0.5524 $-$ 0.4783i)  & (0.5638 $-$ 0.4643i) & (0.5834 $-$ 0.4177i)\\
\hline\hline
\end{tabular*}\label{tableHayward}
\end{table}

 \begin{table}[tbh]
\centering\caption{ QNMs for gravitational perturbations for $\alpha = 0.1, 0.3, 0.4$  for the model of Refs. \cite{Bronnikov}. The parameter $r_0$ is a length scale related to the charge. }
\begin{tabular*}{16.5cm}{*{5}{c @{\extracolsep\fill}}}
\hline \hline
$l$ &n  & $r_0=0.1$   &  $r_0=0.3$ &  $r_0=0.4$ \\ \hline

2 & 0  & 0.3872 $-$ 0.0902i  & 0.4240 $-$ 0.0911i & 0.4499 $-$ 0.0903i\\ \hline

 &   & (0.3877 $-$ 0.0900i)  & (0.4246 $-$ 0.0911i) & (0.4505 $-$ 0.0904i)\\ \hline

2 & 1 & 0.3611 $-$ 0.2774i & 0.4014 $-$ 0.2794i & 0.4298 $-$ 0.2763i\\ \hline

&   & (0.3620 $-$ 0.2766i)  & (0.4027 $-$ 0.2791i) & (0.4316 $-$ 0.2760i)\\ \hline

3 &0 &  0.6213 $-$ 0.0938i & 0.6791 $-$ 0.0949i  & 0.7197 $-$ 0.0942i \\ \hline

&   & (0.6215 $-$ 0.0937i)  & (0.6793 $-$ 0.0949i) & (0.7198 $-$ 0.0942i)\\ \hline

3 &1 &  0.6051 $-$ 0.2844i & 0.6652 $-$ 0.2874i  & 0.7073 $-$ 0.2847i \\ \hline

&   & (0.6054 $-$ 0.2843i)  & (0.6655 $-$ 0.2873i) & (0.7078 $-$ 0.2846i)\\ \hline

3 &2 & 0.5772 $-$ 0.4813i & 0.6409 $-$ 0.4855i & 0.6855 $-$ 0.4802i \\ \hline

&   & (0.5758 $-$ 0.4836i)  & (0.6400 $-$ 0.4871i) & (0.6852 $-$ 0.4812i)\\
\hline\hline
\end{tabular*} \label{tableBronnikov}
\end{table}

 \begin{table}[tbh]
\centering\caption{ QNMs for gravitational perturbations for $\alpha = 0.1, 0.3, 0.4$  for the model of Refs. \cite{Dymnikova}. }
\begin{tabular*}{16.5cm}{*{5}{c @{\extracolsep\fill}}}
\hline \hline
$l$ &n  & $r_0=0.1$   &  $r_0=0.3$ &  $r_0=0.4$ \\ \hline

2 & 0  & 0.3914 $-$ 0.0904i  & 0.4444 $-$ 0.0908i & 0.4888 $-$ 0.0867i\\ \hline

 &   & (0.3919 $-$ 0.0902i)  & (0.4451 $-$ 0.0908i) & (0.4894 $-$ 0.0867i)\\ \hline

2 & 1 & 0.3657 $-$ 0.2772i & 0.4239 $-$ 0.2778i & 0.4714 $-$ 0.2638i\\ \hline

&   & (0.3666 $-$ 0.2773i)  & (0.4255 $-$ 0.2775i) & (0.4736 $-$ 0.2634i)\\ \hline

3 &0 &  0.6279 $-$ 0.0940i & 0.7112 $-$ 0.0946i  & 0.7806 $-$ 0.0904i \\ \hline

&   & (0.6281 $-$ 0.0940i)  & (0.7114 $-$ 0.0946i) & (0.7808 $-$ 0.0904i)\\ \hline

3 &1 &  0.6120 $-$ 0.2850i & 0.6985 $-$ 0.2861i  & 0.7698 $-$ 0.2728i \\ \hline

&   & (0.6123 $-$ 0.2850i)  & (0.6989 $-$ 0.2860i) & (0.7704 $-$ 0.2727i)\\ \hline

3 &2 & 0.5845 $-$ 0.4821i & 0.6762 $-$ 0.4827i & 0.7500 $-$ 0.4587i \\ \hline

&   & (0.5731 $-$ 0.4826i)  & (0.6758 $-$ 0.4839i) & (0.7507 $-$ 0.4589i)\\
\hline\hline
\end{tabular*} \label{tableDymnikova}
\end{table}

 \begin{table}[tbh]
\centering\caption{ QNMs for gravitational perturbations for $q = 0.0, 0.3, 0.6$  for the model of Refs. \cite{ABG}. }
\begin{tabular*}{16.5cm}{*{5}{c @{\extracolsep\fill}}}
\hline \hline
$l$ &n  & $q=0.0$   &  $q=0.3$ &  $q=0.6$ \\ \hline

2 & 0  & 0.3732 $-$ 0.0892i  & 0.3860 $-$ 0.0884i & 0.4438 $-$ 0.0788i\\ \hline

 &   & (0.3737 $-$ 0.0890i)  & (0.3866 $-$ 0.0883i) & (0.4447 $-$ 0.0790i)\\ \hline

2 & 1 & 0.3460 $-$ 0.2749i & 0.3609 $-$ 0.2721i & 0.4240 $-$ 0.2400i\\ \hline

&   & (0.3467 $-$ 0.2739i)  & (0.3623 $-$ 0.2714i) & (0.4275 $-$ 0.2396i)\\ \hline

3 &0 &  0.5993 $-$ 0.0927i & 0.6193 $-$ 0.0921i  & 0.7097 $-$ 0.0826i \\ \hline

&   & (0.5994 $-$ 0.0927i)  & (0.6195 $-$ 0.0921i) & (0.7099 $-$ 0.0826i)\\ \hline

3 &1 &  0.5824 $-$ 0.2814i & 0.6038 $-$ 0.2792i  & 0.6975 $-$ 0.2492i \\ \hline

&   & (0.5826 $-$ 0.2813i)  & (0.6042 $-$ 0.2791i) & (0.6985 $-$ 0.2490i)\\ \hline

3 &2 & 0.5532 $-$ 0.4767i & 0.5771 $-$ 0.4724i & 0.6749 $-$ 0.4191i \\ \hline

&   & (0.5517 $-$ 0.4791i)  & (0.5760 $-$ 0.4745i) & (0.6762 $-$ 0.4189i)\\
\hline\hline
\end{tabular*} \label{tableABG}
\end{table}

\begin{table}[tbh]
\centering\caption{ QNMs for gravitational perturbations for $q = 0.1, 0.3, 0.6$  for the first model of Refs. \cite{new1}. }
\begin{tabular*}{16.5cm}{*{5}{c @{\extracolsep\fill}}}
\hline \hline
$l$ &n  & $q=0.1$   &  $q=0.3$ &  $q=0.6$ \\ \hline

2 & 0  & 0.3738 $-$ 0.0893i  & 0.3791 $-$ 0.0897i & 0.3992 $-$ 0.0909i\\ \hline

 &   & (0.3743 $-$ 0.0890i)  & (0.3796 $-$ 0.0894i) & (0.3997 $-$ 0.0908i)\\ \hline

2 & 1 & 0.3467 $-$ 0.2751i & 0.3524 $-$ 0.2761i & 0.3741 $-$ 0.2796i\\ \hline

&   & (0.3474 $-$ 0.2741i)  & (0.3532 $-$ 0.2752i) & (0.3750 $-$ 0.2789i)\\ \hline

3 &0 &  0.6003 $-$ 0.0928i & 0.6087 $-$ 0.0932i  & 0.6402 $-$ 0.0946i \\ \hline

&   & (0.6005 $-$ 0.0928i)  & (0.6089 $-$ 0.0932i) & (0.6404 $-$ 0.0946i)\\ \hline

3 &1 &  0.5834 $-$ 0.2816i & 0.5921 $-$ 0.2828i  & 0.6246 $-$ 0.2868i \\ \hline

&   & (0.5837 $-$ 0.2814i)  & (0.5923 $-$ 0.2827i) & (0.6249 $-$ 0.2867i)\\ \hline

3 &2 & 0.5543 $-$ 0.4769i & 0.5634 $-$ 0.4787i & 0.5977 $-$ 0.4852i \\ \hline

&   & (0.5528 $-$ 0.4793i)  & (0.5619 $-$ 0.4813i) & (0.5964 $-$ 0.4873i)\\
\hline\hline
\end{tabular*} \label{tablenew1-1}
\end{table}

\begin{table}[tbh]
\centering\caption{ QNMs for gravitational perturbations for $q = 0.1, 0.3, 0.6$  for the second model of Refs. \cite{new1}. }
\begin{tabular*}{16.5cm}{*{5}{c @{\extracolsep\fill}}}
\hline \hline
$l$ &n  & $q=0.1$   &  $q=0.3$ &  $q=0.6$ \\ \hline

2 & 0  & 0.7489 $-$ 0.1786i  & 0.7713 $-$ 0.1801i & 0.8773 $-$ 0.1817i\\ \hline

 &   & (0.7499 $-$ 0.1781i)  & (0.7724 $-$ 0.1797i) & (0.8786 $-$ 0.1817i)\\ \hline

2 & 1 & 0.6948 $-$ 0.5504i & 0.7190 $-$ 0.5544i & 0.8351 $-$ 0.5564i\\ \hline

&   & (0.6962 $-$ 0.5484i)  & (0.7206 $-$ 0.5528i) & (0.8382 $-$ 0.5559i)\\ \hline

3 &0 &  1.2026 $-$ 0.1857i & 1.2379 $-$ 0.1873i  & 1.4043 $-$ 0.1893i \\ \hline

&   & (1.2030 $-$ 0.1856i)  & (1.2382 $-$ 0.1873i) & (1.4046 $-$ 0.1893i)\\ \hline

3 &1 &  1.1689 $-$ 0.5634i & 1.2053 $-$ 0.5682i  & 1.3781 $-$ 0.5729i \\ \hline

&   & (1.1695 $-$ 0.5632i)  & (1.2059 $-$ 0.5680i) & (1.3789 $-$ 0.5728i)\\ \hline

3 &2 & 1.1108 $-$ 0.9544i & 1.1491 $-$ 0.9618i & 1.3326 $-$ 0.9669i \\ \hline

&   & (1.1078 $-$ 0.9592i)  & (1.1463 $-$ 0.9663i) & (1.3315 $-$ 0.9695i)\\
\hline\hline
\end{tabular*} \label{tablenew1-2}
\end{table}

\section{summary}
Although we do not have a complete theory of quantum gravity, regular black hole solutions were proposed by coupling Einstein gravity to an external form of matter. Therefore it is  interesting to compute QNMs for the regular black hole spacetimes and see how it is different from the ordinary ones.
In this work, we have studied  the quasinormal modes of gravitational perturbation around some well-known regular black hole  by using the  WKB approximation and the asymptotic iteration method.  Through numerical calculation, we made a detailed analysis of the gravitational QNM frequencies by varying  the characteristic parameters of the gravitational perturbation and the spacetime charge parameters of the regular black holes. Numerical results show that the imaginary part of quasinormal modes  as  a function of the charge parameter has different  monotonic  behaviors for different black hole spacetimes.  The asymptotic expressions of gravitational QNMs for $l \gg 1$ are computed by using the  eikonal limit method. It is  demonstrated that the gravitational perturbation is  stable in all these spacetimes.

\begin{acknowledgments}
We thank Prof. Ru-Keng Su   for very helpful discussions. We like to thank R.A. Konoplya for providing the WKB approximation. We also like to thank  developers  of the AIM method for their opening codes.  This work is supported partially by the Major State Basic Research Development Program in China (No. 2014CB845402).
\end{acknowledgments}

\begin{figure}[tbp]
\includegraphics[width=13cm,height=19cm]{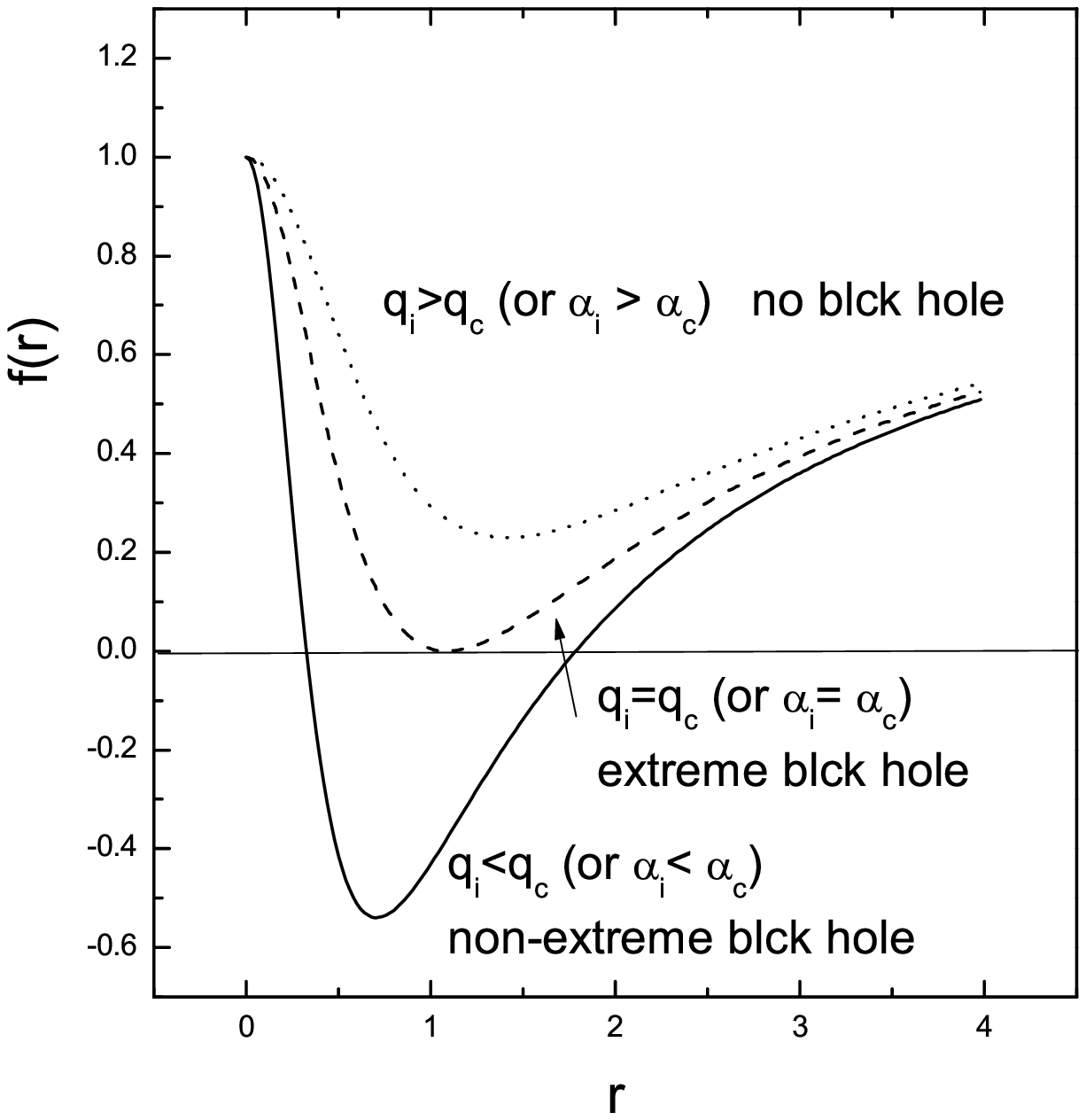}
\caption{Radial dependence of lapse function $f(r)$ of the Bardeen regular black hole for different values of the charge parameter. }
\end{figure}

\begin{figure}[tbp]
\includegraphics[width=13cm,height=19cm]{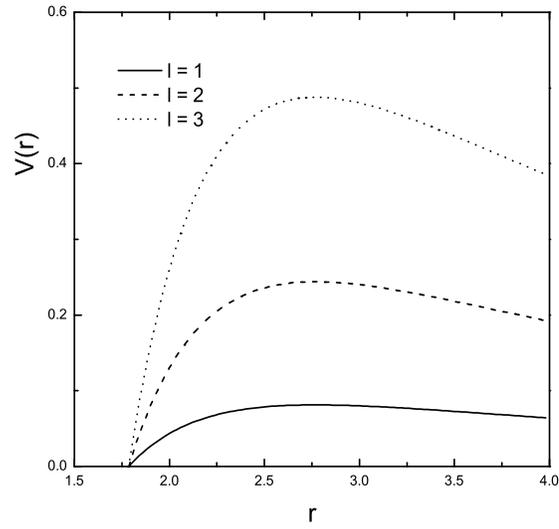}
\caption{Variation of the effective potential $V(r)$ with respect to the radial coordinate $r$ for three values of the angular harmonic index $l$.  }
\end{figure}

\begin{figure}[tbp]
\includegraphics[width=13cm,height=19cm]{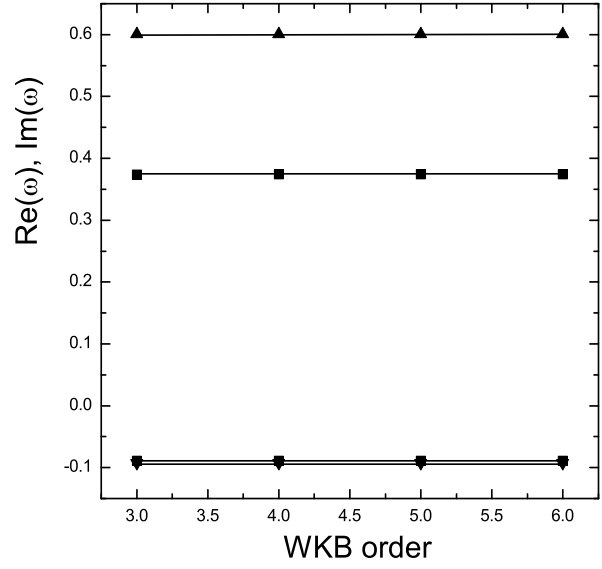}
\caption{ Re($\omega$)  and Im($\omega$) parts of the gravitational QNMs  of the Bardeen black hole with $q=0.1$ as a function of WKB order are given for the $n=0, \;\; l=2$ and $n=0, \;\; l=2$ modes.  }
\end{figure}

\end{document}